\documentstyle[12pt,epsf]{article}
\newcommand{\be}{\begin{equation}}
\newcommand{\ee}{\end{equation}}
\newcommand{\bea}{\begin{eqnarray}}
\newcommand{\eea}{\end{eqnarray}}
\newcommand{\brr}{\begin{array}}
\newcommand{\err}{\end{array}}
\newcommand{\bc}{\begin{center}}
\newcommand{\ec}{\end{center}}
\newcommand{\nn}{\nonumber}

\newcommand{\as}{\alpha_{ s}}

\begin{document}
\pagestyle{empty}
\begin{flushright}
CERN-TH.7283/94 \\
ROME prep. 94/1020 \\
ULB-TH 09/94\\
\end{flushright}
\centerline{\bf{$b \rightarrow s \gamma$ AND
$b \rightarrow s g $: A THEORETICAL REAPPRAISAL}}
\vskip 1cm
\centerline{\bf{ M. Ciuchini$^{a,b}$, E. Franco$^{b}$,
G. Martinelli$^{b,c}$, L. Reina$^{d}$ and L. Silvestrini$^{e}$ }}
\centerline{}
\centerline{}
\centerline{$^a$ INFN, Sezione Sanit\`a, V.le Regina Elena 299,
00161 Roma, Italy. }
\centerline{$^b$ Dip. di Fisica,
Universit\`a degli Studi di Roma ``La Sapienza" and}
\centerline{INFN, Sezione di Roma, P.le A. Moro 2, 00185 Rome, Italy. }
\centerline{$^c$ TH Division, CERN, CH-1211 Geneva 23, Switzerland.}
\centerline{$^d$ Service de Physique Th\'eorique,
Universit\'e Libre de Bruxelles,}
\centerline{Boulevard du Triomphe, CP 225 B-1050 Brussels, Belgium.}
\centerline{$^e$ Dip. di Fisica, Univ. di Roma ``Tor Vergata"
and INFN, Sezione di Roma II,}
\centerline{Via della Ricerca Scientifica 1, I-00173 Rome, Italy.}
\centerline{}
\centerline{}
\begin{abstract}
We present upgraded theoretical predictions for inclusive and exclusive
radiative decays of {\it B} mesons. Our results include those next-to-leading
order corrections that  have  already been computed. Our best
estimates in the Standard Model are $BR(B \rightarrow K^* \gamma)=(4.3
\pm 0.9^{ +1.4}_{-1.0}) \times 10^{-5}$, $BR(B \rightarrow X_s \gamma)
=(1.9 \pm 0.2\pm 0.5) \times 10^{-4}$, $\Gamma(B \rightarrow K^* \gamma)/
 \Gamma(B \rightarrow X_s \gamma)=0.23 \pm 0.09$ and
$BR(B \rightarrow X_s g)= (1.57 \pm 0.15 ^{+0.86}_{-0.59} \pm 0.23)
\times 10^{-3}$. We also consider limits found with
 two-Higgs-doublet models.
\end{abstract}
\vskip 1.5cm
\begin{flushleft}
CERN-TH.7283/94 \\
ROME prep. 94/1020 \\
ULB-TH 09/94\\
June 1994
\end{flushleft}

\newpage
\pagestyle{plain}
\setcounter{page}{1}
\section*{Introduction}
\label{sec:intro}
Radiative decays of {\it B} mesons represent very important tests of the
weak interactions and of the role of effective flavour-changing neutral
currents. Among these
decays, $b \rightarrow s \gamma$ and $b \rightarrow s g$ are theoretically
clean and sensitive to  physics beyond the Standard Model, e.g.
charged scalar Higgs models and/or SUSY models\cite{charged}--\cite{susy}.
The experimental measurement of the exclusive branching fraction
$BR(B \rightarrow K^* \gamma)=(4.5 \pm 1.5 \pm 0.9)\times
10^{-5}$ by the CLEOII
collaboration \cite{cleoii} and the imminent measurement of the inclusive
rate, on which an upper limit
$BR(B \rightarrow X_s \gamma)<5.4\times 10^{-4} $ ($95 \%$
C.L.) \cite{cleoiii}
already exists, offer the opportunity to compare experimental
 results and theoretical  predictions for these quantities.
\par From the theoretical point of view, the prediction of the rates
consists as usual of two steps.
\par  On the one hand, it is necessary to compute the
renormalization of the
coefficients of the effective Hamiltonian to take into account the effects
of strong interactions at short distances, i.e. for scales
$m_b \le  \mu \le M_W$. It turns out that  renormalization effects
 at the leading order (LO) have important consequences,
since they almost double the amplitude  obtained without their
inclusion \cite{shif,wise}.
Unfortunately, the full set of next-to-leading corrections to these decays,
which are necessary for a consistent use of $\Lambda_{QCD}$, of the
renormalization scale, and for more accurate predictions, are not available
yet.
\par  On the other hand, it is necessary to compute
the  hadronic matrix elements of the operators appearing in the effective
Hamiltonian. For inclusive decays, in the framework
of  the Heavy Quark Effective Theory
(HQET), it is possible to predict the rate using the parton model, with
computable corrections
that are expected to be of  order $1/m_b^2$ \cite{inchqet}--\cite{mandrake}.
For exclusive decays, one has to know the relevant hadronic form factors,
obtained from  a non-perturbative estimate.
For $B \rightarrow K^* \gamma$ there is only one form factor,
which  we will denote in  the following  by $F_1(0)$.
  The exclusive channel had a bad reputation
because different predictions of the rate varied
by orders of magnitude. In the recent past, however, lattice
QCD \cite{bhs,ukqcd} and QCD sum
rules \cite{paver}--\cite{narison} have procured more reliable results and
the theoretical uncertainties have been substantially reduced.
This makes
the exclusive decays more interesting as tests of the Standard Model.
\par We present an upgraded analysis of the inclusive and exclusive
$b \rightarrow s \gamma$  and
of the inclusive $b \rightarrow s g$ decay rates, which takes into account
 several improvements  made recently:
\begin{itemize}
\item The coefficient of the magnetic (chromo-magnetic) operator,
$C_7$ ($C_8$) is now established.
In refs. \cite{silve1,silve2}, the long-standing problem of the
regularization dependence of the LO coefficients of these operators
was solved; the results were  found to be different from all previous
calculations \cite{altri}. Those of refs. \cite{silve1,silve2} were
subsequently confirmed in ref. \cite{curci}.
\item The next-to-leading order corrections to the anomalous dimension
matrix are partially known \cite{buras,noi}. We will make use of this
information to try to evaluate the effect of the next-to-leading
terms and include it in the error on the final  predictions.
We also compare results obtained in the HV and NDR regularization
schemes, since a spurious regularization dependence is introduced
by the incomplete NLO terms. The NLO corrections that we can already
include diminish $C_7$  in both the HV and NDR schemes.
It turns out that in HV the dependence on the renormalization scale
is substantially reduced. A similar effect is also found  in NDR,
where nevertheless a sizeable  dependence is still present. We will also
make use of the partial calculation of the $O(\as)$ corrections to the
inclusive rate, which has been computed in ref. \cite{ali}.
\item We will study both the inclusive and exclusive channels,
the latter being ignored in most of more recent analyses.
We will make use of lattice and QCD sum rules predictions for
the relevant form factor $F_1(0)$.
\item By varying within their errors the experimental and theoretical
quantities, we obtain a distribution of values for the theoretical
predictions, from which we estimate the theoretical uncertainty.
\end{itemize}
\par Our predictions  contain several
differences with respect to the  recent analysis of ref. \cite{munz}:
\par i) the  estimates we present are also
for the exclusive $B \rightarrow K^* \gamma$ rate;
\par ii) according to refs. \cite{beneke,bigi}, we have used the running mass,
and not the pole mass, for the evaluation of the inclusive rate;
\par iii) $O(1/m_b^2)$ terms have been taken into account in the present study;
\par iv) NLO corrections to the coefficient function \cite{ali} and
to the anomalous dimension matrix \cite{buras,noi} have been included by us
in  evaluation of the values and theoretical uncertainties for the rates.
\par Of all these differences, the most important is the last one,
since the known NLO corrections diminish the strong $\mu$ dependence
and reduce the values of the rates. Apart from this,
 the bulk of our results is substantially in agreement with those
of ref. \cite{munz}.

\section*{\bf Basic Formulae}
The effective Hamiltonian responsible for
$b \rightarrow s \gamma$ and
$b \rightarrow s g $ decays can be written as \cite{shif,wise},
\cite{silve1}--\cite{curci}:
\be H^{eff}= - V_{tb} V^*_{ts} \frac{G_F}{\sqrt{2}}
\left( C^{eff}_7(\mu) O_7(\mu) + C^{eff}_8(\mu) O_8(\mu) \right)
, \nn \ee
where:
\bea O_7 &=& \frac{Q_d e}{16\pi^2} m_b \left( \bar s \sigma^{\mu\nu}
b \right)_R F_{\mu\nu} \nn \\
O_8 &=& \frac{ g_s}{16\pi^2} m_b \left( \bar s \sigma^{\mu\nu}
t^A b \right)_R G^A_{\mu\nu} \nn \eea
and  $\mu$ is the renormalization scale of the relevant operators.
{}From the effective Hamiltonian one can derive the decay rates.
In the following, we report the main formulae that will be used in the
numerical evaluation of the inclusive and exclusive branching fractions.
\par {\bf Inclusive $B \rightarrow X_s \gamma$}:
\be BR(B \rightarrow X_s \gamma )= \left[ \frac{\Gamma(
B \rightarrow X_s \gamma)}{\Gamma(B \rightarrow X l \nu_l)} \right]^
{th} \times BR^{exp} (B \rightarrow X l \nu_l), \label{eq:bri} \ee
where : \be \left[ \frac{\Gamma(
B \rightarrow X_s \gamma)}{\Gamma(B \rightarrow X l \nu_l)} \right]^
{th} =
\frac{\vert V_{ts}^* V_{tb} \vert^2}{\vert V_{cb}\vert^2} \frac{\alpha_e}
{6 \pi g(m_c/m_b)} \times F \times \vert C_7^{eff}(\mu) \vert ^2
\label{eq:par} \ee
with the phase-space factor $g(z)$ given by:
\be g(z)=1-8 z^2 +8 z^6-z^8-24 z^4 \ln(z) \nn \ee
and
\be F= \frac{K(m_t/M_W,\mu)}{\Omega(m_c/m_b,\mu)} \label{factor}, \nn \ee
In eq. (\ref{factor}) the quantity $\Omega(z)$ contains the $O(\alpha_s)$
QCD corrections to the semileptonic decay rate \cite{maia,kuhn}. Within a good
approximation it is given by \cite{sirlin}:
\be
\Omega(z,\mu)\simeq 1-\frac{2\alpha_s(\mu)}{3\pi}\left[\left(\pi^2-
\frac{31}{4}\right)(1-z)^2+\frac{3}{2}\right];
\label{kappa} \nn \ee
the factor $K(m_t/M_w,\mu)$ contains the $O(\as)$ NLO corrections
to the $B \rightarrow X_s \gamma$ rate, due to
real and virtual gluon emission \cite{ali}. The calculation of
ref. \cite{ali} does not contain the full set of
 $O(\as)$ corrections, which would
require a two-loop calculation of the coefficient function and a three
loop calculation of the anomalous dimension. In ref. \cite{ali},
the authors only included
some terms, which one may argue to be the most important ones. We have
used the results of ref. \cite{ali} in our analysis,
 in the same spirit in which we considered the
 NLO anomalous dimension, which is only partially known.
 \par
The scale $\mu$, which appears in some of the  previous formulae, is there for
two reasons. In the numerator of eq. (\ref{eq:par}), $\mu$ denotes the
renormalization scale  of the effective $b \rightarrow s \gamma$
Hamiltonian. It also represents the scale at which we decided to
compute the expansion parameter $\as$ for  the  QCD corrections to
the semileptonic decay rate. For simplicity we have taken the same
value of  $\mu$ in
both cases. According to refs. \cite{beneke,bigi}, in the framework of the
HQET, to avoid problems with renormalon (non-perturbative) effects, we
have used as expansion parameter the running mass $m_b=m_b(\mu)$.
For this reason, since the $b \rightarrow s \gamma$ operators are
evaluated at a generic $\mu$, we have omitted the  factor
$\left( m_b(\mu)/
m_b({\rm pole})\right)^2$ which was included in
$F$  by the authors of ref. \cite{munz},
see eq. (\ref{factor}).
\par {\bf  Exclusive $B \rightarrow K^* \gamma$}:
\bea
BR(B \rightarrow K^* \gamma )\!\!&=&\!\!
\left[ \frac{\Gamma(B \rightarrow K^* \gamma)}
{\Gamma(B \rightarrow X_s\gamma)} \right]^{th}\times
\left[ \frac{\Gamma(B \rightarrow X_s \gamma)}
{\Gamma(B \rightarrow X l \nu_l)} \right]^{th} \nn\\
\!\!&\times&\!\! BR^{exp} (B \rightarrow X l \nu_l)
, \label{eq:bre}
\eea
where
\be
\left[ \frac{\Gamma(B \rightarrow K^* \gamma)}
{\Gamma(B \rightarrow X_s \gamma)} \right]^{th}=
\left(\frac{M_b}{m_b}\right)^3\left(1-\frac{M_{K^*}^2}{M_B^2}\right)^3
 \times \frac{1}{1 +
(\lambda_1-9 \lambda_2)/(2 m_b^2)}\times \vert F_1(0) \vert^2
\label{yyy} \ee
In the HQET formalism, the parameters $\lambda_1$ and $\lambda_2$
describe the leading non-perturbative corrections (at order $O(1/m_b^2)$)
to the parton model predictions for the inclusive rate  \cite{mandrake}.
They are related to the kinetic energy of the {\it b} quark (inside the
{\it B} meson) and to the $B$--$B^*$ mass splitting.
The $1/m_b^2$ corrections cancel in
$\Gamma(B \rightarrow X_s \gamma)/
\Gamma(B \rightarrow X l \nu_l)$ but not in the ratio (\ref{yyy});
$F_1(0)$ is the relevant form factor defined by:
\be \langle K^*(p^\prime, \eta) \vert \bar s \sigma_{\mu\nu}
(1 + \gamma_5) q^\nu b \vert B(p) \rangle  =
 2 i \epsilon_{\mu\nu\rho\sigma} \eta^{* \nu}p^\rho p^{\prime \sigma}
F_1(q^2)\nn \ee \be  + 2 \left[ \eta^*_\mu (M_B^2 -M_{K^*}^2) -
(\eta^* \cdot q ) (p+p^\prime)_\mu \right] G_2(q^2) \nn \ee
with $G_2(0)=F_1(0)/2$.
\par {\bf Inclusive $B \rightarrow X_s g$}:
\par The theoretical estimate of  $BR(B \rightarrow X_s g)$ is at present
much rougher than $BR(B \rightarrow X_s \gamma)$. Even
a partial evaluation of the coefficient function
and $1/m_b^2$ terms is missing in this case.
We give our  prediction, which improves with respect to  previous calculations
only because $C^{eff}_8$ has been corrected and we include
 the NLO terms of the $(6 \times 6)$ anomalous-dimension submatrix.
To evaluate the branching ratio we have used the formula:
\be BR(B \rightarrow X_s g )= \left[ \frac{\Gamma(
B \rightarrow X_s g)}{\Gamma(B \rightarrow X l \nu_l)} \right]^
{th} \times BR^{exp} (B \rightarrow X l \nu_l), \label{eq:brig} \ee
where : \be \left[ \frac{\Gamma(
B \rightarrow X_s g)}{\Gamma(B \rightarrow X l \nu_l)} \right]^
{th} =
\frac{\vert V_{ts}^* V_{tb} \vert^2}{\vert V_{cb}\vert^2} \frac{2
\alpha_s(\mu)}
{ \pi g(m_c/m_b)} \frac{1}{\Omega(m_c/m_b,\mu)}
  \times \vert C_8^{eff}(\mu) \vert ^2
. \label{eq:parg} \ee
\vfill
\section*{{\bf Predictions and Uncertainties \newline
for the Decay Rates}}

\begin{table}[t]
\begin{center}
\begin{tabular}{||c|c||}
\hline
Parameters & Values \\ \hline\hline
$\vert V_{ts}^* V_{tb} \vert^2/\vert V_{cb}\vert^2$ & $0.95 \pm 0.04$ \\
\hline
$m_c/m_b$ & $0.316\pm 0.013$ \\ \hline
$m_t$(GeV) & $174 \pm 16$ \\ \hline
$\lambda_1$(GeV$^2$) & $-0.15\pm 0.15$ \\ \hline
$\lambda_2$(GeV$^2$) & $0.12 \pm 0.01$ \\ \hline
$m_b(\mu=m_b)$(GeV) & $4.65 \pm 0.15$ \\ \hline
$F_1(0)$ & $ 0.35 \pm 0.05$ \\ \hline
$BR^{exp} (B \rightarrow X l \nu_l)$ & $0.107 \pm 0.005$ \\ \hline
$\Lambda_{QCD}^{n_f=5}$(GeV) & $0.240 \pm 0.090$ \\ \hline
$\mu$ & $m_b/2$--$2 m_b$ \\ \hline
\end{tabular}
\caption[]{ \it{Values of the different parameters used to predict
the inclusive and exclusive radiative b decay rates. }}
\label{tab:qua}
\end{center}
\end{table}

We give in table \ref{tab:qua} the range of variation of
all  the quantities appearing  in eqs. (\ref{eq:bri})--(\ref{yyy}), which
have been used to obtain our results.
\par
\begin{figure}[c]   
    \begin{center}
       \setlength{\unitlength}{1truecm}
       \begin{picture}(6.0,6.0)
          \put(-6.0,-6.2){\special{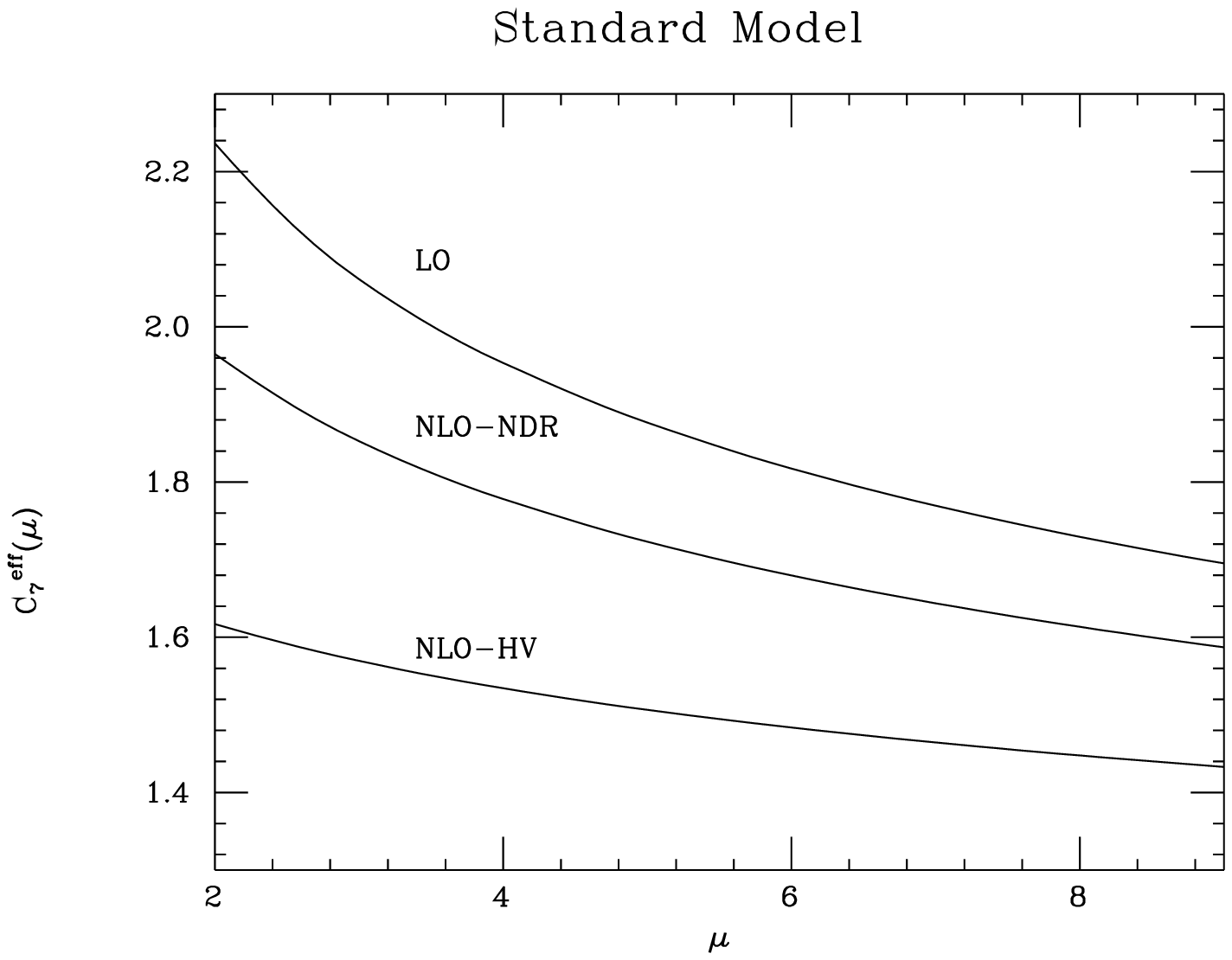}}

       \end{picture}
    \end{center}
       \caption[]{\it{$C_7^{eff}(\mu)$ as a function of $\mu$
in the LO, NLO-HV and NLO-NDR cases. For the definition of NLO see the text.}}
    \protect\label{fig:mudep}
\end{figure}

\begin{figure}[tb]   
    \begin{center}
    \epsfysize=14truecm
    \leavevmode\epsffile{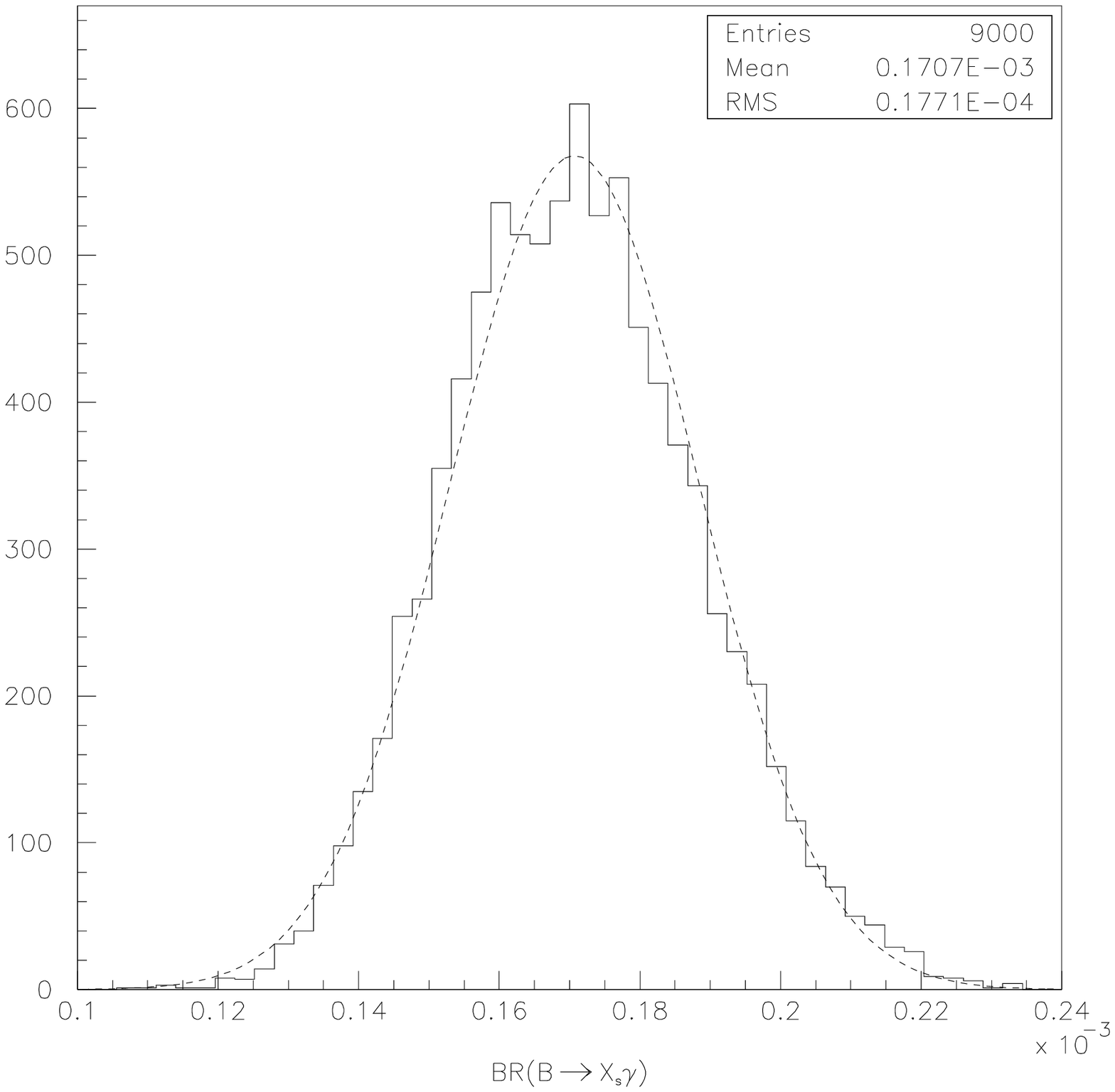}
%
    \end{center}
       \caption[]{\it{Distribution of values for
$BR(B \rightarrow X_s \gamma)$ for $\mu=m_b$.}}
    \protect\label{fig:dis}
\end{figure}

 Using the central values of table \ref{tab:qua},
 we show in fig. \ref{fig:mudep}
 the $\mu$ dependence of $C_7^{eff}(\mu)$
at  the LO and NLO. For LO we mean that we have taken the
anomalous dimension matrix at the LO. For the total  rate,
we  put to zero all the $O(\as)$
terms that appear in eqs. (\ref{eq:bri})--(\ref{yyy}), including those
relative  to the semileptonic rate. In this case one has  also to use
$K(m_t/M_W,\mu)=1$. In the NLO case we turn on all the known NLO
corrections, including the $(6 \times 6)$ anomalous-dimension matrix
computed in refs. \cite{buras,noi}. In this case we have varied
$K(m_t/M_W,\mu)$ according to the results of ref. \cite{ali}, i.e.
$0.79 \le K(m_t/M_W,\mu) \le 0.86$ for $m_b/2\le  \mu \le 2 m_b$.
We notice that the $\mu$ dependence is reduced both in the HV and
in the NDR cases.
If we call $R=(C_7^{eff}(\mu)\vert_{\mu=m_b/2}/
C_7^{eff}(\mu )\vert_{\mu=2 m_b})^2$, we get
$R^{LO} \sim 1.72$, $R^{NDR} \sim 1.54$ and $R^{HV}
\sim 1.25$ \footnote{This means that the coefficient itself varies only
of $\sim 10-15 \%$ in a range of scales as large as $\mu \sim 2-9$
GeV.}.
One would prefer $HV$ because of the reduced $\mu$ dependence,
as shown fig. \ref{fig:mudep}.
In the  absence of a complete calculation, however,
 we cannot decide which of the
results, NDR or HV, is closer to the  real NLO result. For this reason,
for all the quantities reported below, we will combine the results
obtained with NDR and HV with $m_b/2 \le \mu \le 2 m_b$,  and include the
differences in the  final estimate of the error. As central value
we will take the average between the NDR and HV result.
Regarding all the other quantities, we allow them to vary within the ranges
reported in table \ref{tab:qua}, with Gaussian distributions
for the experimental parameters and flat distributions for the theoretical
ones. In a given regularization scheme and for a fixed value of
 $\mu$, this procedure generates a pseudo-Gaussian
distribution of values (see fig. \ref{fig:dis})
from which we deduce the error. For more details,
see  ref. \cite{epenew}.

\begin{table}[t]
\begin{center}
\begin{tabular}{||c|c|c|c||}
\hline
$\phantom{\mu}$ & \multicolumn{3}{|c||}{$BR(B \rightarrow X_s \gamma)\times
10^4$} \\ \hline

$\mu$ (GeV) & LO & NLO$_{HV}$ & NLO$_{NDR}$ \\ \hline

$m_b/2$ & $3.81\pm 0.47$ & $1.92\pm 0.19$ & $2.77\pm 0.32$ \\ \hline

$m_b$ & $2.93\pm 0.33$ & $ 1.71\pm 0.18$ & $2.25\pm 0.25$ \\ \hline

$2 m_b$ & $2.30\pm 0.26$ & $ 1.56\pm 0.17$ & $1.91\pm 0.21$ \\ \hline \hline

$\phantom{\mu}$ & \multicolumn{3}{|c||}{$BR(B \rightarrow K^* \gamma)\times
10^5$} \\ \hline

$\mu$(GeV) & LO & NLO$_{HV}$ & NLO$_{NDR}$ \\ \hline

$m_b/2$ & $6.9\pm 1.5$ & $4.4\pm 0.8$ & $6.4\pm 1.3$ \\ \hline

$m_b$ & $5.3\pm 1.1$ & $3.8\pm 0.8$ & $5.0\pm 1.0$ \\ \hline

$2 m_b$ & $4.2\pm 0.9$ & $ 3.3\pm 0.7$ & $4.1\pm 0.8$ \\ \hline \hline

$\phantom{\mu}$ & \multicolumn{3}{|c||}{$BR(B \rightarrow X_s g)\times 10^3$}
\\ \hline

$\mu$(GeV) & LO & NLO$_{HV}$ & NLO$_{NDR}$ \\ \hline

$m_b/2$ & $4.21\pm 0.38\pm 1.03$ & $1.81\pm 0.17\pm 0.14$ &
$3.25\pm 0.30\pm 0.66$ \\ \hline

$m_b$ & $2.66\pm 0.24\pm 0.47$ & $ 1.45\pm 0.14\pm 0.10$ &
$2.19\pm 0.21\pm 0.32$ \\ \hline

$2 m_b$ & $1.81\pm 0.17\pm 0.25$ & $ 1.19\pm 0.12\pm 0.09$ &
$1.58\pm 0.15\pm 0.19$ \\
\hline
\end{tabular}
\caption[]{ {\it Theoretical predictions for exclusive and
inclusive radiative B decays.  The last error on
$BR(B \rightarrow X_s g)$ comes from $\Lambda_{QCD}$. We have preferred to
show it separately, since the rate is directly proportional to $\as$.}}
\label{tab:br}
\end{center}
\end{table}

We want to add some comments on the value and error
of $F_1(0)$ which has been used in the exclusive case.
Recent lattice
and QCD sum rules calculations of this quantity give:
$F_1(0)=0.20 \pm 0.02 \pm 0.06$ \cite{bhs},
 $F_1(0)=0.30^{+10}_{-8}$ \cite{ukqcd} and
$F_1(0)=0.35 \pm 0.05 $ \cite{paver}, $F_1(0)=0.32 \pm 0.05$ \cite{ali2},
$F_1(0)=0.310 \pm 0.013 \pm 0.033 \pm 0.060 $ \cite{narison}.
For this reason we have chosen to use
$F_1(0)=0.35 \pm 0.05$, which covers most of the theoretical predictions
(see table \ref{tab:qua}).
{}From the numbers given in table \ref{tab:br} we quote
$BR(B \rightarrow K^* \gamma) = (4.3
\pm 0.9^{ +1.4}_{-1.0}) \times 10^{-5}$,
$ BR(B \rightarrow X_s \gamma)
=(1.9 \pm 0.2\pm 0.5) \times 10^{-4}$
 and $\Gamma(B \rightarrow K^* \gamma)/
 \Gamma(B \rightarrow X_s \gamma)=0.23 \pm 0.09$.
The first error comes from
the width of the pseudo-Gaussian distribution of the theoretical values,
the second includes the $\mu$ dependence and regularization dependence
of the results.
\par The results for $BR(B \rightarrow X_s g )$
are reported in table \ref{tab:br}. Because of the factor
$\as(\mu)$ present in eq. (\ref{eq:parg}), the dependence on $\Lambda_{QCD}$
is stronger than for $BR(B \rightarrow X_s \gamma )$, and we have reported
the corresponding  uncertainty separately in the table.
Our best estimate is
$  BR(B \rightarrow X_s g ) = (1.57 \pm 0.15 ^{+0.86}_{-0.59} \pm 0.23)
\times 10^{-3} $,
where the last error is due to $\Lambda_{QCD}$.
\par
\section*{Constraints on 2H models \newline
from radiative $B$ decays}
\begin{figure}[t]   
    \begin{center}
    \epsfysize=14truecm
    \leavevmode\epsffile{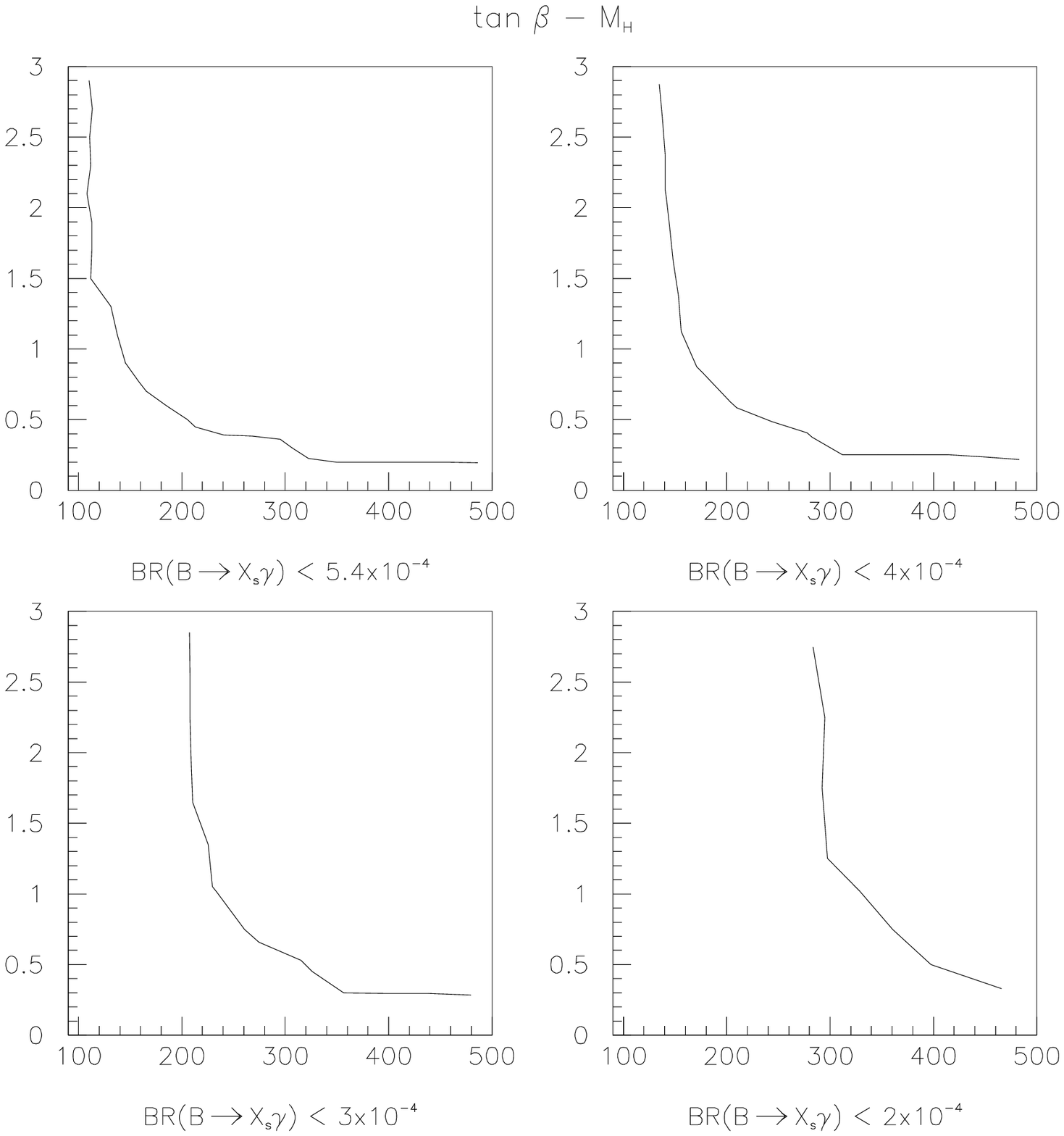}
    \end{center}
       \caption[]{\it{Exclusion regions (on the left of
the figures) in the plane  $M_{H^+}-
\tan\beta$, at next-to-leading order, for different
upper bounds of $BR(B \rightarrow X_s \gamma)$.}}
    \protect\label{fig:exclu}
\end{figure}

\begin{figure}[t]   
    \begin{center}
       \setlength{\unitlength}{1truecm}
       \begin{picture}(6.0,6.0)
          \put(-6.0,-6.2){\special{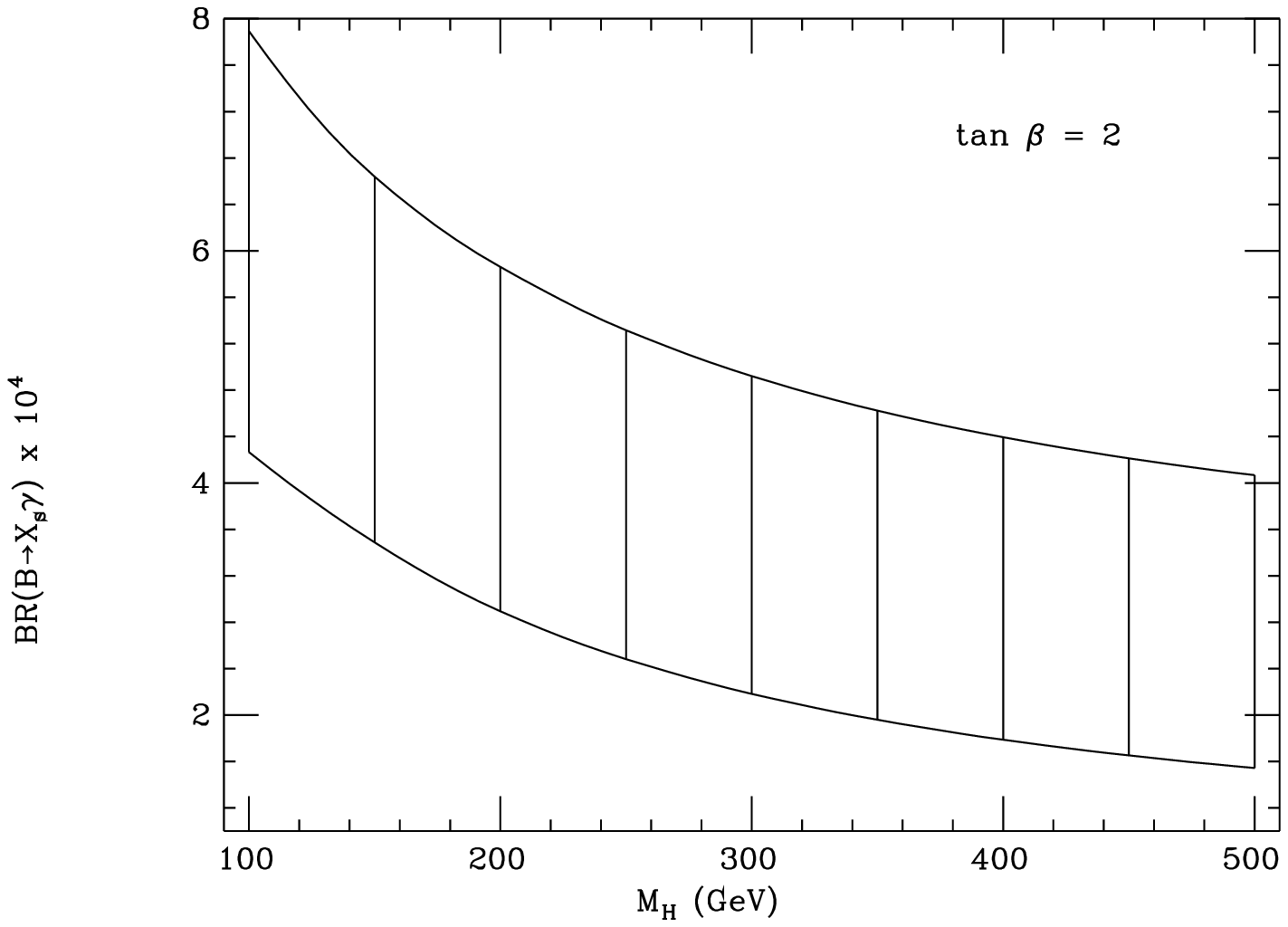}}

       \end{picture}
     \end{center}
       \caption[]{\it{Predictions for  $BR(B \rightarrow X_s \gamma)$
in the 2H model with $\tan \beta=2$ are given as a function of
$M_{H^+}$.}}
    \protect\label{fig:band}
\end{figure}

We consider the popular 2H model known  in the literature
 as Model II \cite{charged}. We look for bounds in the plane  $M_{H^+}-
\tan\beta$, by imposing the experimental constraints on the branching
ratios for $B\to X_s\gamma$ and $B\to K^*\gamma$. We
modify the initial conditions of $C^{eff}_7(M_W)$
 to take into account the charged Higgs
contributions \cite{condi} and
calculate eqs. (\ref{eq:bri})--(\ref{eq:par}) and (\ref{eq:bre})--(\ref{yyy}),
by varying the parameters with the same criteria as before.
In presence of charged Higgs,
the corrections of $O(\as)$ to $C_i(M_W)$, $i=1,\dots,6$
  are known \cite{condias}
 and were included at the NLO  in the present calculation.
 We look for bounds  by  varying
the parameters $M_{H^+}$,$\tan\beta$ in the range $M_{H^+} \ge 91
\,\, {\rm GeV}$,
$\tan\beta \le 0.48 \,\, {\rm GeV}^{-1}\times M_{H^+}$
\cite{tbeta}--\cite{diemoz}.
The first condition comes from
the lower bound on the Higgs mass established  from the limits on
Higgs production at LEP. The second is obtained from the comparison between
theoretical calculations \cite{tbeta} and the experimental
measurements of
$BR^{exp} (B \rightarrow X \tau \nu_\tau)$ \cite{diemoz}.
Since the exclusion regions are substantially independent of $\tan \beta$
for large values of this parameter, the results will only be given
for $\tan \beta \le 3$.
We observe that the exclusion region is strongly reduced  by going from the
LO to the NLO.
This is mainly due to the sizeable  reduction of the rate when the
known  NLO corrections are considered.  Because of the larger theoretical
error,
the exclusive channel cannot be used to put a limit.
The results are given in fig. \ref{fig:exclu}.
 In view of future measurements of $BR(B \rightarrow X_s \gamma)$,  and
biased by our theoretical predictions,
we also give the exclusion region corresponding to  an upper limit
for  $BR(B \rightarrow X_s \gamma)$ of  $ 4 \times 10^{-4},
 3 \times 10^{-4}$, and $ 2 \times 10^{-4}$. Finally,  in fig. \ref{fig:band},
$BR(B \rightarrow X_s \gamma)$ is reported as a function of $M_{H^+}$ for
$\tan \beta=2$. In the figure, the  band indicates the theoretical
uncertainty.
\section*{{\bf Acknowledgements}}
We warmly thank A. Ali, I. Bigi, A. Buras, M. Crisafulli, M. Diemoz, T.
Mannel, G. Ridolfi, M. Serone, M. Shifman
and N.G. Uraltsev for very useful discussions.
We acknowledge the partial support of the MURST, Italy, and  the INFN.

\newpage
%

%

\end{document}